\documentclass{article}

\usepackage{arxiv}

\usepackage{babel}
\usepackage[utf8]{inputenc} 
\usepackage[T1]{fontenc}    
\usepackage{hyperref}       
\usepackage{url}            
\usepackage{booktabs}       
\usepackage{amsfonts}       
\usepackage{nicefrac}       
\usepackage{microtype}      
\usepackage{lipsum}
\usepackage{graphicx}
\graphicspath{ {./Images/} }

\usepackage{microtype}   
\usepackage{doi}
\usepackage{amsmath,amssymb,amsfonts,amsbsy}
\usepackage{algorithmic}
\usepackage{graphicx}
\usepackage{textcomp}
\usepackage{xcolor}
\usepackage{url}
\usepackage{multicol}
\usepackage{newtxtext,newtxmath}
\usepackage{soul}
\usepackage{float}
\usepackage{subfigure} 
\usepackage{adjustbox}
\usepackage{comment}
\usepackage{float}
\usepackage{siunitx}
\usepackage{cite}
\usepackage{multirow}
\usepackage{array} 
\usepackage{tikz}
\usetikzlibrary{shapes,arrows,shadows,positioning}
\usetikzlibrary{matrix,chains,positioning,decorations.pathreplacing}
\newcommand{\wh}[1]{\widehat{#1}}
\newcommand{\bs}[1]{\boldsymbol{#1}}

\title{Hybrid Pipeline SWD Detection in Long-Term EEG Signals}

\author{
    Antonio~Quintero-Rinc\'on, Nicolas Masino, Verónica Marsico\\
    Department of Data Science, Data Science and AI Laboratory.\\ Catholic University of Argentina (UCA), Argentina, Buenos Aires, Argentina \\
    \texttt{antonioquintero@uca.edu.ar}
\And
    Hadj~Batatia\\
    MACS School, Heriot-Watt University, Dubai Campus, United Arab Emirates.\\
\And
    To cite this work, please use this reference:\\
    SABI2025/CLIC2025, 136, pp. 1–14, 2025. \\
    \doi{10.1007/978-3-032-06401-1\_90}}

\begin{document}
\maketitle
\begin{abstract}
Spike--and--wave discharges (SWDs) are the electroencephalographic hallmark of absence epilepsy, yet their manual identification in multi-day recordings remains labour-intensive and error-prone. We present a lightweight hybrid pipeline that couples analytical features with a shallow artificial neural network (ANN) for accurate, patient-specific SWD detection in long-term, monopolar EEG. A two-sided moving-average (MA) filter first suppresses the high-frequency components of normal background activity. The residual signal is then summarised by the mean and the standard deviation of its normally distributed samples, yielding a compact, two-dimensional feature vector for every \SI{20}{\second} window. These features are fed to a single-hidden-layer ANN trained via back-propagation to classify each window as \emph{SWD} or \emph{non-SWD}. The method was evaluated on \num{780} channels sampled at \SI{256}{\hertz} from 12 patients, comprising 392 annotated SWD events.  
It correctly detected 384 events (sensitivity: \SI{98}{\percent}) while achieving a specificity of \SI{96.2}{\percent} and an overall accuracy of \SI{97.2}{\percent}.  
Because feature extraction is analytic, and the classifier is small, the pipeline runs in real-time and requires no manual threshold tuning.
These results indicate that normal-distribution descriptors combined with a modest ANN provide an effective and computationally inexpensive solution for automated SWD screening in extended EEG recordings.
\end{abstract}


\section{Introduction}

Electroencephalography (EEG) is a non-invasive technique that records the brain’s electrical activity through scalp electrodes.  
Because it is inexpensive, portable, tolerant of subject movement, and offers millisecond-level temporal resolution \cite{Hamalainen1993,Regan2010}, EEG remains the frontline tool for clinical neurophysiology and, in particular, for the diagnosis and monitoring of epilepsy.

Epilepsy is the most prevalent chronic neurological disorder after migraine, affecting an estimated \SI{0.5}{\percent}–\SI{1}{\percent} of the global population \cite{EpilepsySpectrum2012,WHO2024Epilepsy}.  
Clinical manifestations range from brief, focal motor disturbances to generalised convulsions and loss of consciousness.  
At the neuronal level, seizures correspond to paroxysmal, rhythmic discharges produced by large ensembles of synchronised neurons.  
These discharges give rise to characteristic epileptiform patterns, spikes, polyspikes, sharp waves, and spike-and-waves. \emph{spike-and-waves discharges} (SWDs) refer to episodes or bursts of the spike-and-wave pattern, whose morphology, frequency, and spatial distribution inform both diagnosis and therapeutic planning. They describe the occurrence of spike-and-wave activity over a period of time.

Reliable seizure detection is challenging. Long-term EEG traces are replete with transient artefacts, including eye blinks, muscle activity, electrode pops, and sleep phenomena, that can mimic epileptiform transients and lead to false alarms \cite{NiedermeyerDaSilva2010,QuinteroRincon2021}.  
Although computer-aided seizure-detection methods (SDMs) have steadily improved, expert review of multi-day recordings is still labour-intensive and error-prone, underscoring the need for robust automatic tools.

SWDs are highly regular, symmetrical patterns that typify absence epilepsy and are also encountered in Lennox–Gastaut and Ohtahara syndromes \cite{Markand1977,NiedermeyerDaSilva2010,Quintero2018b}.  
Animal models, especially rodents, have greatly advanced the understanding of SWDs \cite{Pinault2001,VanHese2009,Ovchinnikov2010,Bergstrom2013,Rodgers2015,Blumenfeld2005,Avoli2012}.  
Nevertheless, human studies remain relatively scarce and often rely on manual annotation \cite{QuinteroRincon2020a,Luttjohann2022,Gobbo2021,QuinteroRincon2019d,Samie2018}.

Artificial-neural-network (ANN) approaches have been explored for automated seizure detection since the mid-1990s \cite{Webber1996,Wilson2004}.  
Recent work spans feed-forward, convolutional, and recurrent architectures \cite{Handa2023,Zhang2023,QuinteroRincon2019b,Turk2019,Sriraam2018,Assi2018,Hussein2018,Truong2018,Acharya2018,Schirrmeister2017,Johansen2016}, combined with diverse feature sets such as spikes and sharp waves \cite{Gabor1992}, high-voltage spike-and-wave spindles \cite{Jando1993}, and bivariate connectivity measures.  
Training strategies include classical back-propagation \cite{Gevins1986,Guo2010,Dhif2017}, weight adaptation for early detection \cite{Aguiar2015}, self-organising maps \cite{Gabor1996}, and Lyapunov-based stability analysis \cite{Ponce2016}.

Despite these advances, deep models are not always practical in resource-constrained clinical settings, and their interpretability can be limited.  
Motivated by these concerns, we propose a lightweight hybrid pipeline that couples a simple analytical descriptor with a conventional ANN for SWD detection in long-term EEG.  
The pipeline summarises each \SI{20}{\second} window by the mean ($\mu$) and the standard deviation ($\sigma$) of its normally distributed samples after a two-sided moving-average (MA) filter; these two parameters form a compact input vector for ANN classification.  
This design revisits classical statistical descriptors in a modern framework, demonstrating that low-dimensional feature spaces can still yield high sensitivity and specificity while remaining interpretable and computationally inexpensive.

The remainder of the paper is organised as follows. Section~\ref{sec:mat} details the dataset, the MA-based feature extraction, and the ANN model. Experimental results are presented in Section~\ref{sec:res}. Finally, Section~\ref{sec:con} concludes and outlines future work.


\section{Materials and Methods}
\label{sec:mat}

\subsection{Database}
\label{ssec:data}
\begin{figure}[!ht]
	\centering
	\subfigure[SWD]{\includegraphics[width=30mm]{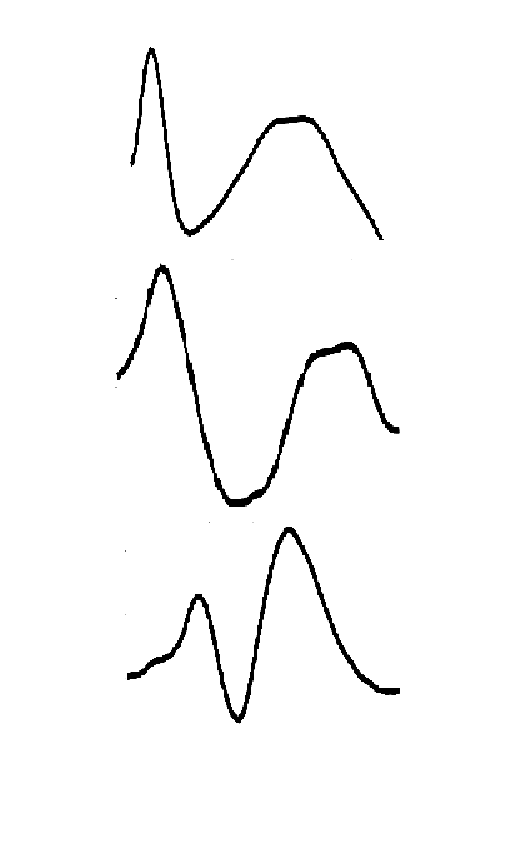}}
    \hspace*{0.5em}
	\subfigure[EEG raw]{\includegraphics[width=120mm]{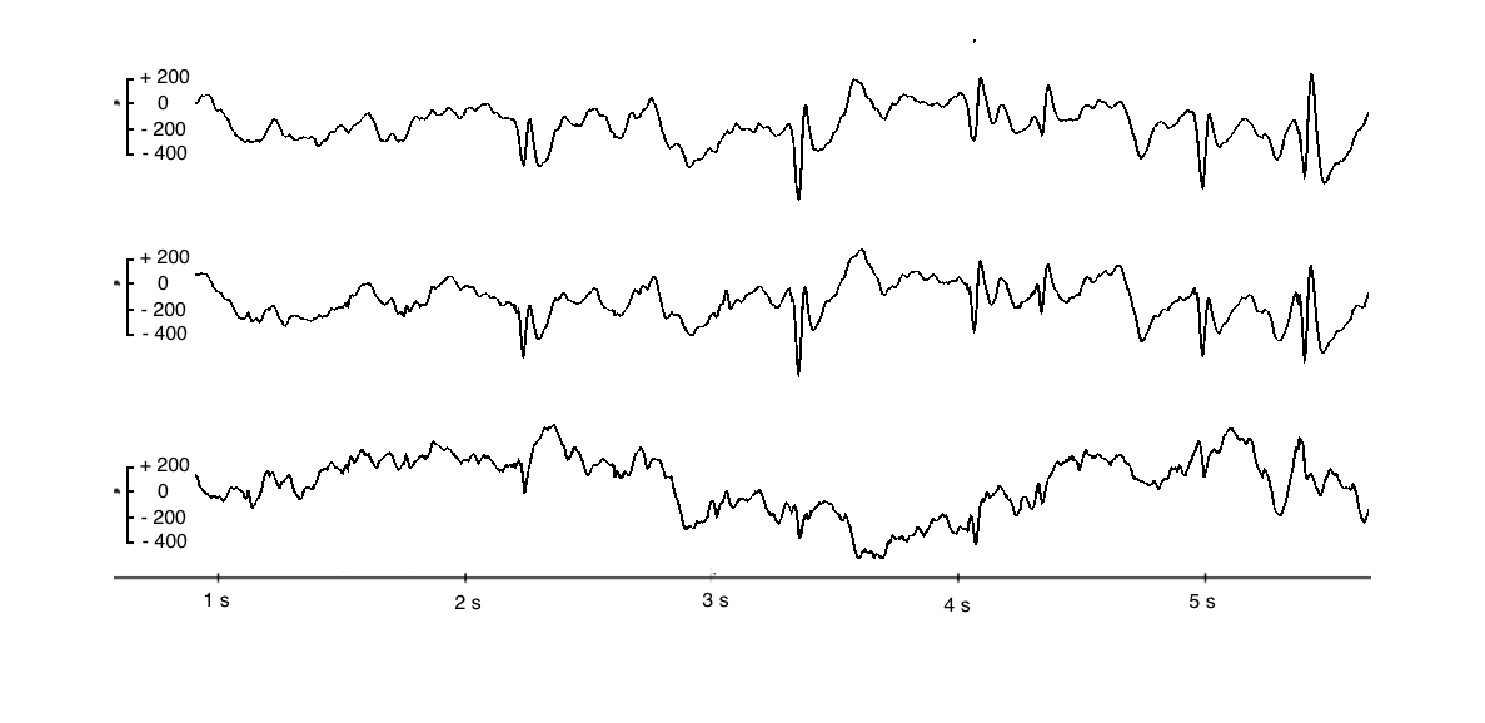}}
	\caption{(a) Symmetric and regular morphology of three spike-and-wave discharges (SWD) and (b) three channels of EEG raw with the patterns in a clinical environment.}
	\label{fig:raw}
\end{figure}

A database with 780 balanced monopolar 256 Hz signals was created by an expert neurologist specializing in epilepsy: 390 spike-and-wave discharges, denominated by SWD, and 390 non-spikes-and-wave signals denominated by nSWD, measured from 12 different patients from Fundación Lucha contra las Enfermedades Neurológicas Infantiles (FLENI). A standard 10-20 EEG system was used, featuring 22 channels: Fp1, Fp2, F7, F3, Fz, F4, F8, T3, C3, Cz, C4, T4, T5, P3, Pz, P4, T6, O1, O2, Oz, FT10, and FT9. The EEGs were recorded in a routine clinical setting during sleep with six patients diagnosed with Refractory or drug-resistant epilepsy (DRE) and six with focal epilepsy, ages ranging from 4 to 40 years. The hospital's ethics committee approved the study. Recordings included non-seizure activities and artifacts such as head and body movements, chewing, blinking, early sleep stages, and electrode pops or displacements. A cascade of a second-order low-pass filter at 30 Hz and a first-order high-pass filter at 1600 Hz was used in the software, a common approach in EEG processing to improve signal quality and facilitate neural activity analysis.
The spike-and-wave discharge signals have different times and waveforms, but their morphology is preserved, while the non-spike-and-wave signals have normal waveforms, see Fig. \ref{fig:raw} (a) and (b).  See \cite{QuinteroRincon2020a} for more details about this database.

\subsection{Moving Average}
\label{ssec:ma}
The Moving Average (MA) is a technique extensively used to smooth out short-term variability and highlight more enduring trends or cycles \cite{Hyndman2021}. In the present study, MA is proposed to enhance the structural distinction between SWD and nSWD as follows.

Let \(\{X_t\}_{t=1}^{n}\) be the EEG time series with samples \(x_1,x_2,\dots,x_n\).  
For a two-sided moving average, choose an integer half-window size \(h\ge 0\).  
The window centred at index \(t\) then contains \(2h+1\) consecutive samples  
\(x_{t-h},\dots,x_{t},\dots,x_{t+h}\).  
The moving-average (MA) value is

\begin{equation}
    m_t \;=\; \frac{1}{2h+1}\sum_{j=-h}^{h} x_{t+j},
    \quad
    t = h+1,\;h+2,\;\dots,\;n-h .
    \label{eq:ma}
\end{equation}

Consequently, \(\{M_t\} = \{m_{h+1},\dots,m_{\,n-h}\}\) is the smoothed series obtained from $\{X_t\}$.  
If the final window would extend beyond the available data (\(t+h>n\)), reduce \(h\) accordingly to the last position or truncate the series so that every average is computed over a complete window.

Because an SWD is typified by a protracted wave succeeding a spike, an order $k$, the moving average is not capable of accurately estimating both the spike and the wave concurrently. Nevertheless, it is reasonable to assume that the difference between two moving averages of differing orders enhances the variability of $\{X_t\}$. 
For a high-frequency time series such as nSWD, the two-sided moving average reduces the variability of normal events because it spans a broader temporal context. By contrast, spike-and-wave patterns contain longer low-frequency components and therefore do not reduce variability to the same extent \cite{QuinteroRincon2020a}.

Let $\wh{M}_1$ and $\wh{M}_2$ be the two-sided moving averages, with $k_1$ and $k_2$ successive points, respectively, where $k_1<k_2$. 
Consequently, the disparity between these two moving averages can be expressed as Eq.~(\ref{eq:equ3}).  
\begin{align}
    \wh{P} = \wh{M}_1-\wh{M}_2
\label{eq:equ3}    
\end{align}

\subsection{Statistical Modelling}
\label{sec:sig}
The probability density function (PDF) $f_\textnormal{n}$ of the normal distribution is given by
\begin{align}
\label{eq:PDF}
f_\textnormal{n}(p |\mu,\sigma^2) = \frac{1}{\sigma\sqrt{2\pi}}\exp^{-\frac{1}{2} \left( \frac{p-\mu}{\sigma}\right)^2}
\end{align}
where $-\infty < \mu < \infty$ is the location (mean) parameter, $\sigma>0$ is the  scale (standard deviation) parameter, and $P$ is the moving average difference from $X$ (see Eq.~\eqref{eq:equ3}). 

We consider that spike-and-wave discharge signals are distributed according to a normal distribution with parameters $\bs \theta_{1}=(\mu_1,\sigma_1)$. On the other hand, non-spike-and-wave signals follow a normal distribution with parameters $\bs \theta_{0}=(\mu_0,\sigma_0)$.
This work is based on the hypothesis that the parameter vectors $\bs \theta_{1}$ and $\bs \theta_{0}$ are characteristics of the two corresponding signals. Therefore, they can be used to distinguish spike-and-wave episodes within long-term EEG signals.
To show the correctness of this method, the two parameters $(\mu,\sigma)$ were estimated from each signal in our $780$-sample dataset (Section \ref{ssec:data}). 

A maximum likelihood estimation method has been used for this purpose \cite{Thomopoulos2017}, which consists of minimizing the log-likelihood $\mathcal{L}(\mu, \sigma^2)$: 
\begin{align}
\label{eq:like}
\mathcal{L}(\mu, \sigma^2)&= \sum^n_{i=1}\log(f_n) \\
\nonumber
&=-\sum^n_{i=1}\log(\sigma\sqrt{2\pi})-\sum^n_{i=1}\frac{1}{2}\left(\frac{p_i-\mu}{\sigma}\right)^2
\end{align}
Differencing $\mathcal{L}(\mu, \sigma^2)$ for each parameter yields:
\begin{align}
    \frac{\partial \mathcal{L}(\mu,\sigma^2)}{\partial \mu} &= \frac{1}{\sigma^2}\sum^n_{i=1}(p_i-\mu) =0 \\
    \frac{\partial \mathcal{L}(\mu, \sigma^2)}{\partial\sigma^2} &= -\frac{n}{2\sigma^2}+\frac{1}{2\sigma^4}\sum^n_{i=1}(p_i-\mu)^2=0
\end{align}
Then, the MLE unbias $\wh{\mu}$ parameter and the bias $\wh{\sigma}^2$ parameter are estimated as:
\begin{align}
    \wh{\mu} &=\frac{1}{n}\sum^n_{i=1}p_i \\
    \wh{\sigma}^2 &= \frac{1}{n}\sum^n_{i=1}(p_i-\wh{\mu})^2\\
    \wh{\sigma} &= \sqrt{\wh{\sigma}^2} 
\end{align}

\subsection{Artificial Neural Network Model}
\label{ssec:net}
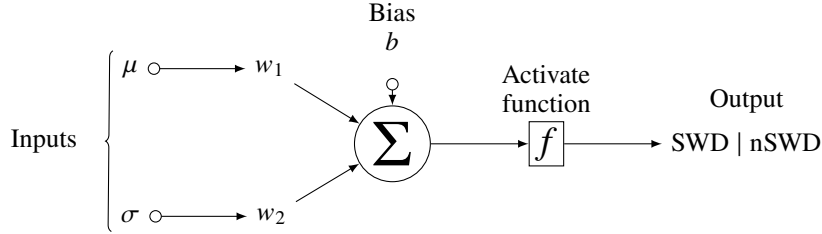
\begin{figure}[H]
\centering
\begin{tikzpicture}[
	init/.style={
		draw,
		circle,
		inner sep=2pt,
		font=\Huge,
	},
	squa/.style={
		draw,
		inner sep=2pt,
		font=\Large,
		join = by -latex
	},
	start chain=2,node distance=13mm
	]
	\node[on chain=2]
	(dummy) {};
    \node[on chain=2](w3){};
	\node[on chain=2,init] (sigma) 
	{$\displaystyle\Sigma$};
	\node[on chain=2,squa,label=above:{\parbox{2cm}{\centering Activate \\ function}}]   
	{$f$};
	\node[on chain=2,label=above:Output,join=by -latex] 
	{\centering SWD~$|$~nSWD};

	\begin{scope}[start chain=1]
	\node[on chain=1] at (0,1cm) 
	(x1) {$\mu$};
	\node[on chain=1,join=by o-latex] 
	(w1) {$w_1$};
	\end{scope}

	\begin{scope}[start chain=3]
	\node[on chain=3] at (0,1-1cm) 
	(x2) {$\sigma$};
        \node[on chain=3,join=by o-latex]
        (w2) {$w_2$};
	\end{scope}
    
	\node[label=above:\parbox{2cm}{\centering Bias \\ $b$}] at (sigma|-w1) (b) {};

	\draw[-latex] (w1) -- (sigma);
	\draw[-latex] (w2) -- (sigma);
	\draw[o-latex] (b) -- (sigma);
	
	\draw[decorate,decoration={brace,mirror}] (x1.north west) -- node[left=10pt] {Inputs} (x2.south west);	
\end{tikzpicture}
\caption{Structure of the artificial neural network used in the proposed methodology. Note that, $\mu$ and $\sigma$ are the normal parameters}
\label{fig:ann}
\end{figure}

To recognise spike-and-wave events, we supplied the normal-distribution parameters to a feed-forward ANN with a hidden layer of $n$ neurons. The network was trained with back-propagation to classify each event as either spike-and-wave or non-spike-and-wave. \\
For the signal $S_i$, let $(p_{1},p_{2}) = (\mu_i,\sigma^2_i)$ the input vector with the properties of EEG signal, then
\begin{align}
\bs	\delta &=f(\bs \eta) = \frac{1}{1+ \exp^{-\eta}}, \\
\bs \eta &= \sum_{i=1}^{m} w_{i}p_{i}
\end{align}
where $\bs \delta$ is the output class: (1) for spike-and-wave and (0) for non-spike-and-wave; $\bs \eta = \bs W^{T} \bs X$ is the dot scalar product of input at each neuron, $\bs W = [w_{1},w_{2},\cdots, w_{m}]^T$ is a weight vector, and $f(.)$ is the transfer sigmoid function, see Fig. \ref{fig:ann}.

For a supervised network, data consists of $n$ training couples $\{(\bs \theta_{1}, c_{1}), \cdots, (\bs \theta_{n}, c_{n})\}$, where $\bs \theta$ is the normalized input vector and $c$ is the normalized desired output. The goal of the training stage is to minimize a cost function $\lambda(w)$ using the backpropagation function overall $n$ pattern vectors. Usually, $\lambda(w)$ is the squared error.
\begin{align}
\lambda(w) = \sum_{p=1}^{n}\frac{(\delta_{p} - \bs \theta_{p})^2}{n}
\end{align}
We refer the reader to \cite{NNDesign2014} for a comprehensive treatment of the design and properties of neural networks.

\section{Results and discussion}
\label{sec:res}
In this section, we evaluate the proposed methodology using our dataset of  $780$ EEG monopolar $256$ Hz signals, with $390$ SWD and 390 nSWD, from 12 different patients presented above in subsection \ref{ssec:data}. 

\begin{figure}[!ht]
	\centering
	\subfigure[Normal distribution parameters before MA]{\includegraphics[width=100mm]{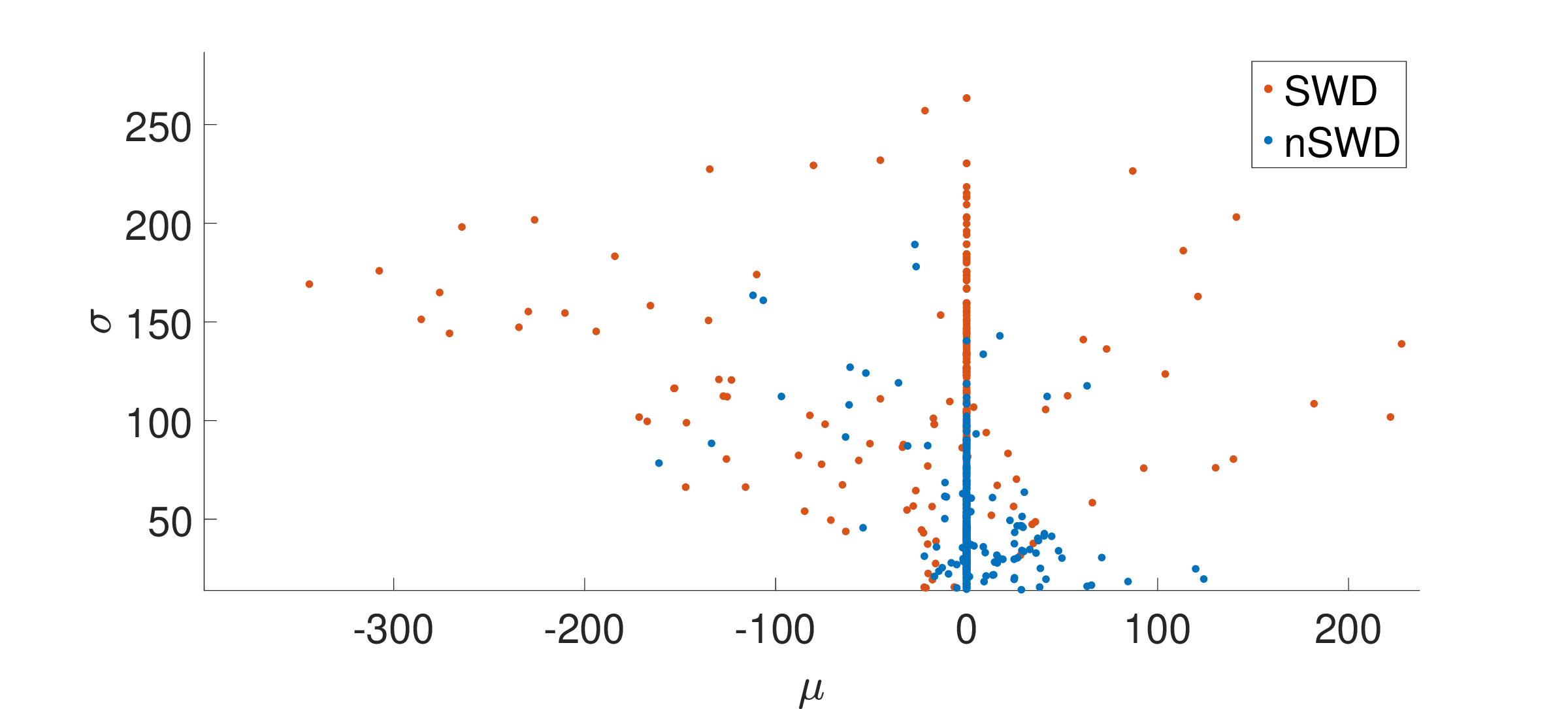}}
    \hspace*{0.5em}
	\subfigure[Normal distribution parameters after MA]{\includegraphics[width=100mm]{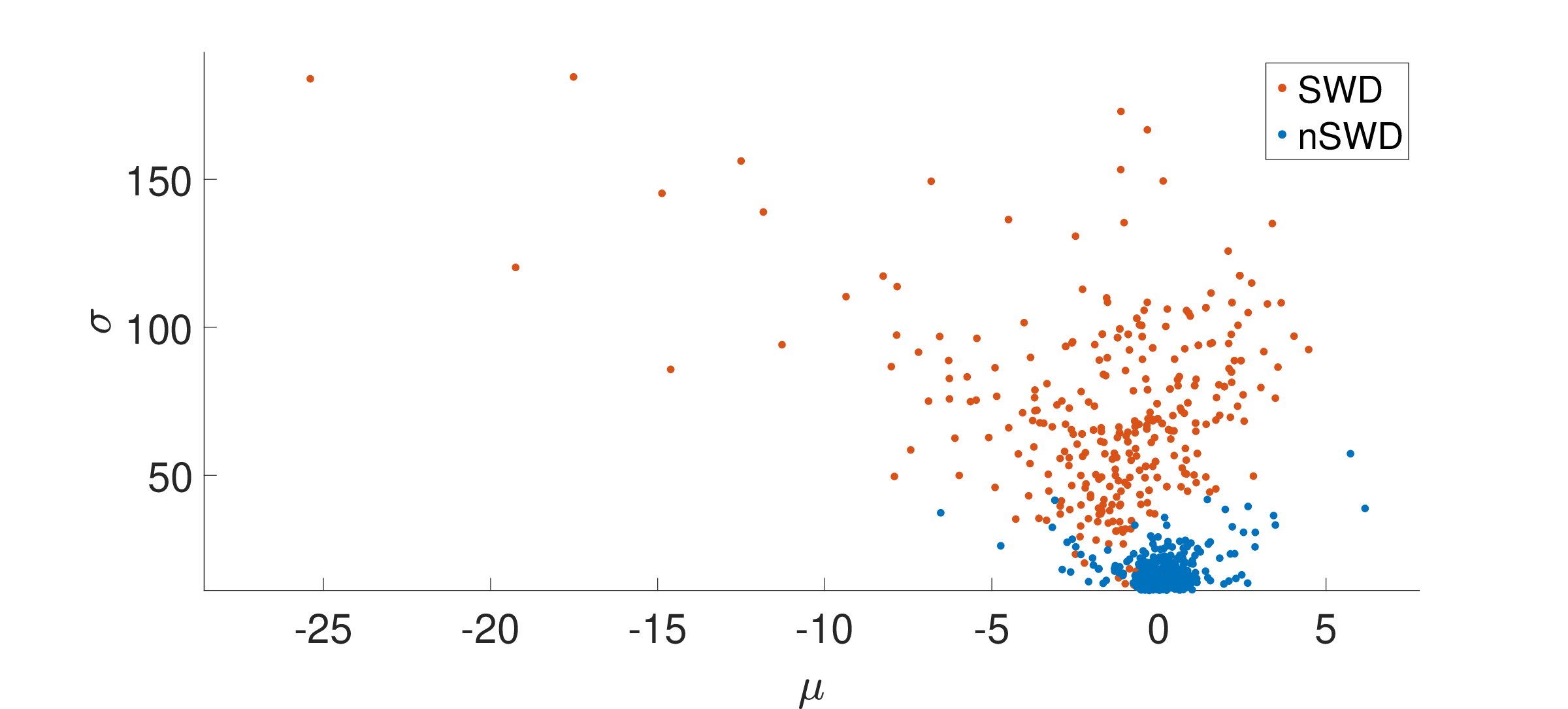}}
	\caption{Scatter plots for the normal parameters $\mu$ and $\sigma$ representing the spike-and-wave events (orange dots) and non-spike-and-wave events (blue dots) are presented. It is evident that SWD events typically exhibit a higher scale $\sigma$, whereas nSWD events generally display lower values for the location $\mu$ and scale $\sigma$ parameters both before and after the application of the MA decomposition.}
\label{fig:n}
\end{figure}

Fig.~\ref{fig:n} shows the scatter plots of the estimated parameters before and after applying MA, with orange dots representing SWD events and blue dots representing nSWD events. Before applying MA, the SWD events tend to exhibit a higher scale $\sigma$, whereas nSWD events show a location $\mu$ and scale $\sigma$ with low values. However, after applying MA, the distinction between the two events is entirely clear. SWD events exhibit a higher scale $\sigma$, whereas nSWD events exhibit a location $\mu$ and scale $\sigma$ with low values. Notably, the behavior of the $\sigma$ scale parameter is useful for correctly discriminating between different events. 

Table \ref{tab:stats} confirms quantitatively that the scale parameter $\sigma$ is of interest, as shown visually in Fig. \ref{fig:n}. 
This finding is consistent with previous reports \cite{QuinteroRincon2018a}. In those studies, signal variance, modelled by a generalised Gaussian distribution, was shown to be closely related to the variability of brain activity.
Notice that the means and standard deviations differ markedly between the two event types. These differences enable a simple threshold-based detection scheme \cite{QuinteroRincon2021}.

\begin{table}[!ht]
\centering
\begin{tabular}{|c| c| c|| c| c||}
\hline
& $\mu_{\text{SWD}}$ & $\mu_{\text{nSWD}}$ & $\sigma_{\text{SWD}}$ & $\sigma_{\text{nSWD}}$ \\ 
\hline\hline
mean Before	MA& -11.8788 & 1.4355 & 117.9236 & 46.2041  \\
\hline
std Before 	MA&   4520.6 & 438.2142 & 2428.1 & 731.9147 \\
\hline\hline
mean After	MA& -1.3629 & 0.0867 & 71.1112 & 15.3159 \\
\hline
std After MA&   10.8607 & 0.9981 & 866.3190 & 49.5937 
\\
\hline

\end{tabular}
\vspace{0.3cm}
\caption{Means and Standard Deviations of the Normal Distribution Parameters: location ($\mu$) and scale ($\sigma$) for spike-and-wave discharges (SWD) and non-spike-and-wave discharges (nSWD) before and after applying MA.}
\label{tab:stats}
\end{table}

Table~\ref{tab:effectsize} delineates the features of Cohen's d quantification alongside the $p$-values for the estimated parameters both before and after the application of the moving average (MA). Cohen's d serves as a standardized effect size metric that assesses the disparity between the means of two classes by standard deviation units. An increase in the absolute value of Cohen's d signifies that the feature possesses a more robust discriminative capacity \cite{MasinoQuintero2024,Masino_Quintero-Rincon_2025}. Notably, the variation between the parameters of the normal distribution before and after its moving average implementation indicates an improvement in effect size. This finding suggests that the method effectively enhances the detection of spike-and-wave activities in EEG signals, which is supported by the low $p$-values.

\begin{table}[!ht]
\centering
\begin{tabular}{|c| c| c|| c| c||}
\hline
& $\mu$ & $\mu_{\text{$p$-value}}$ & $\sigma$ & $\sigma_{\text{$p$-value}}$ \\ 
\hline\hline
Cohen's d Before MA& -0.28283 & 9.6596e-05 & 1.8695 & 3.4290e-107  \\
\hline
Cohen's d After	MA&  -0.63008 & 1.5219e-17 & 2.7718 & 7.6789e-182 
\\
\hline
\end{tabular}
\vspace{0.3cm}
\caption{Cohen's d and $p$-values for the estimated parameters before and after applying MA.}
\label{tab:effectsize}
\end{table}

\begin{figure}[!ht]
\centering
\vspace*{-1em}
\includegraphics[width=100mm]{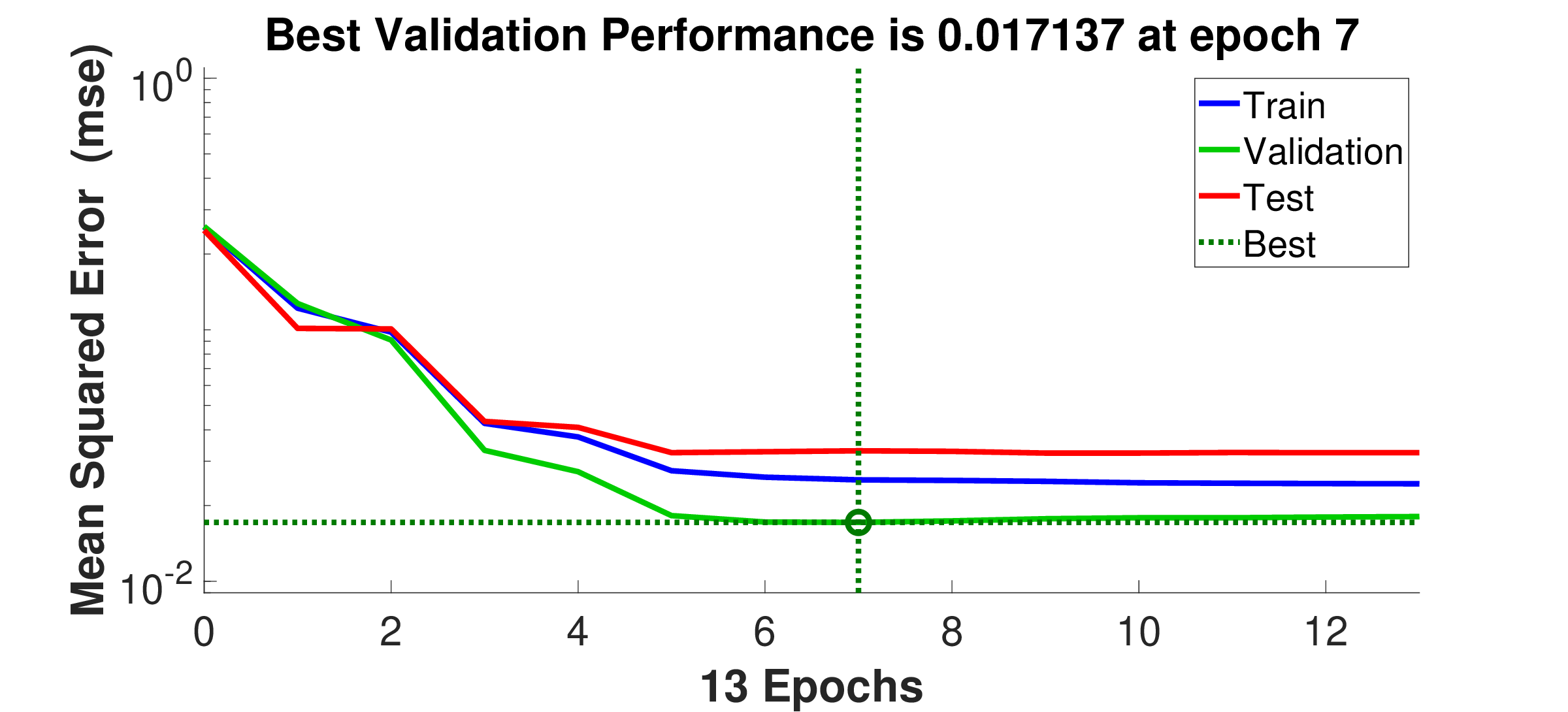}
\caption{Artificial neural network fit performance for training (blue color), validation (green color), and testing  (red color) stages. Note that all signals decay until epoch 5.}
\label{fig:perf}
\end{figure}

\begin{figure}[!ht]
\centering
\includegraphics[width=100mm]{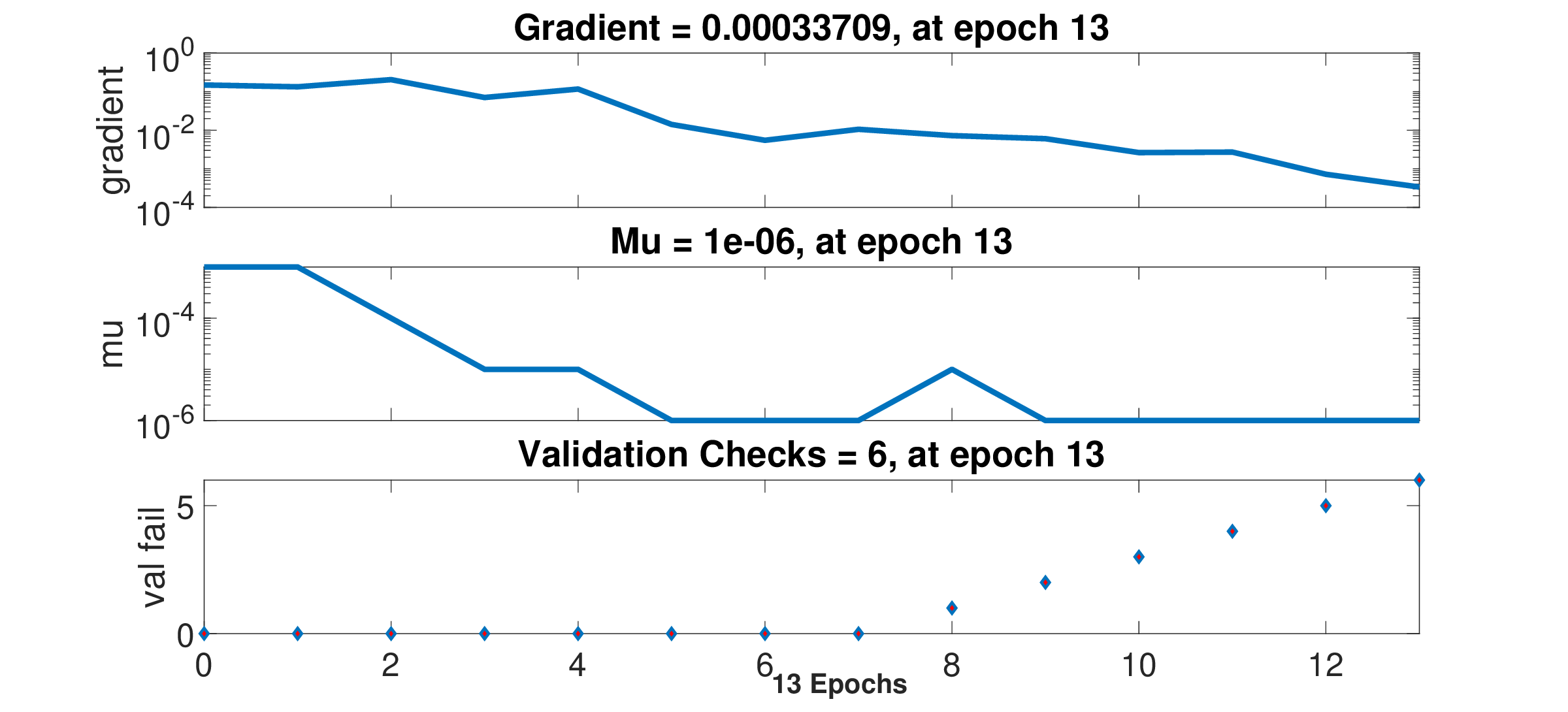}
\caption{Training state shows low values of the gradient, which suggests that our model is a powerful tool for detecting spike-and-wave discharges.}
\label{fig:trainstage}
\end{figure}

For the classification stage, the signals were distributed randomly as follows: 70\% of the signals were destined for the training stage (blue color), $15\%$ for the validation stage (green color), and $15\%$ for the testing stage (red color). The hidden layer size was defined empirically according to input parameters from the normal distribution. Fig.~\ref{fig:perf} shows the performance fit using the normal parameters, all signals decay until epoch 5 with $\lambda=0.027$ for train signals, $\lambda=0.018$ for validation signals, and $\lambda=0.032$ for test signals. The best performance arises in epoch 7 with $\lambda=0.025$ for training signals, $\lambda=0.017$ for validation signals, and $\lambda=0.033$ for test signals. The three signals remain constant after epoch 6 until epoch 13, with $\lambda=0.025$ for the training signals, $\lambda=0.017$ for the validation signals, and $\lambda=0.032$ for the test signals.
Therefore, a very low $\lambda$ at the end of the training stage suggests a well-trained artificial neural network model \cite{Twomey1995}, see Fig. \ref{fig:trainstage}. Small $\lambda$ values, close to zero, indicate that the desired outputs and the Neural Network's outputs for the training set have become very similar; this is a good training model for pattern recognition. This can be verified with the low values of the gradient and $\mu$ in Fig. \ref{fig:trainstage}. Table \ref{tab:epochs} summarizes the main epoch values for our model.

\begin{figure}[!ht]
\centering
\includegraphics[width=100mm]{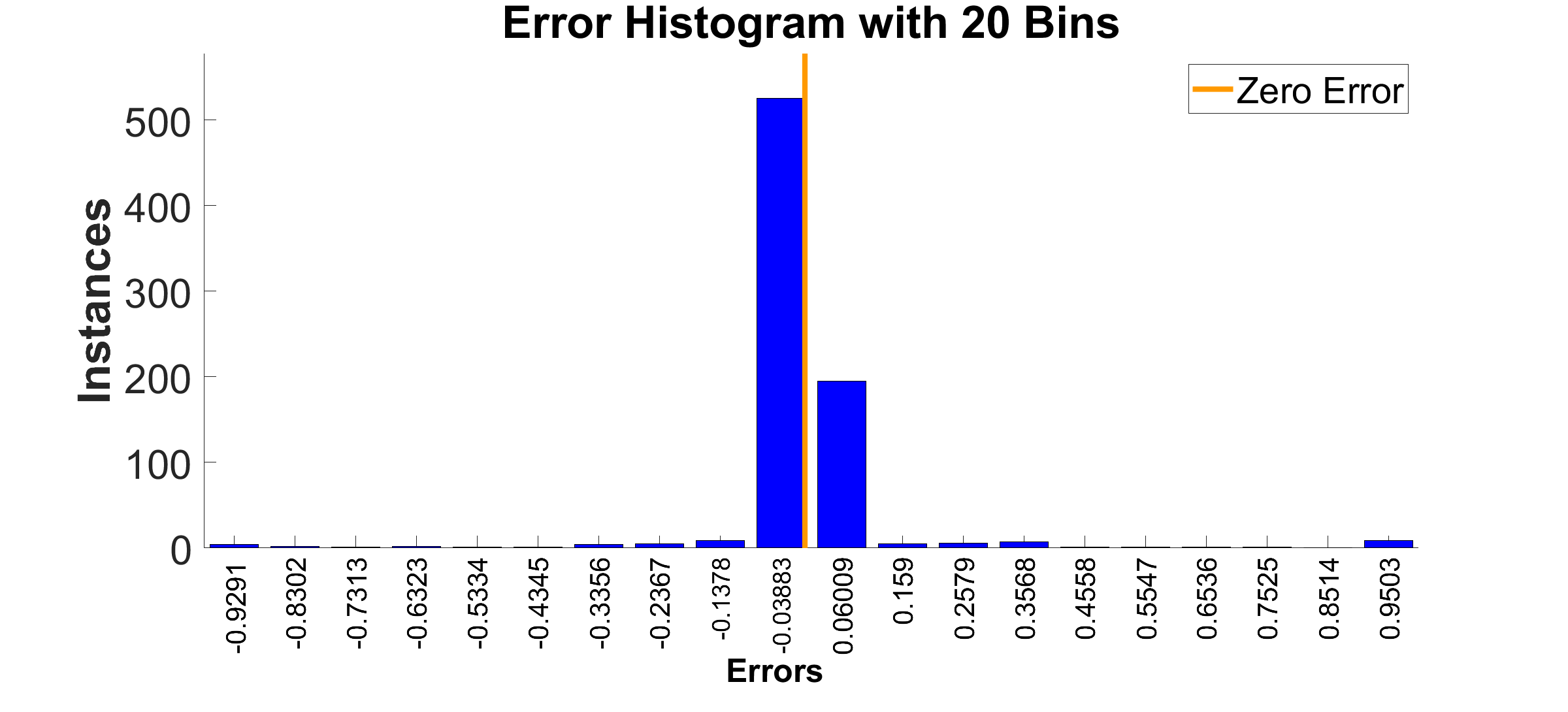}
\caption{Error histogram. The tail presents high values at the ends of the shape, while the small values are close to zero, where the highest density of the form is concentrated. }
\label{fig:histog}
\end{figure}

Fig.~\ref{fig:histog} shows the error histogram for the training, validation, and testing stages for $20$ bins. Note that the error values are defined as the difference between targets and outputs. For the normal parameters, the tail presents high values at the ends of the shape, while the small values are close to zero, where the highest density of the form is concentrated. This shows consistent performance measures that can be corroborated in Fig.~\ref{fig:perf} and Fig.~\ref{fig:trainstage} with the low values of the gradient.

\begin{figure}[!ht]
\centering
\includegraphics[width=75mm]{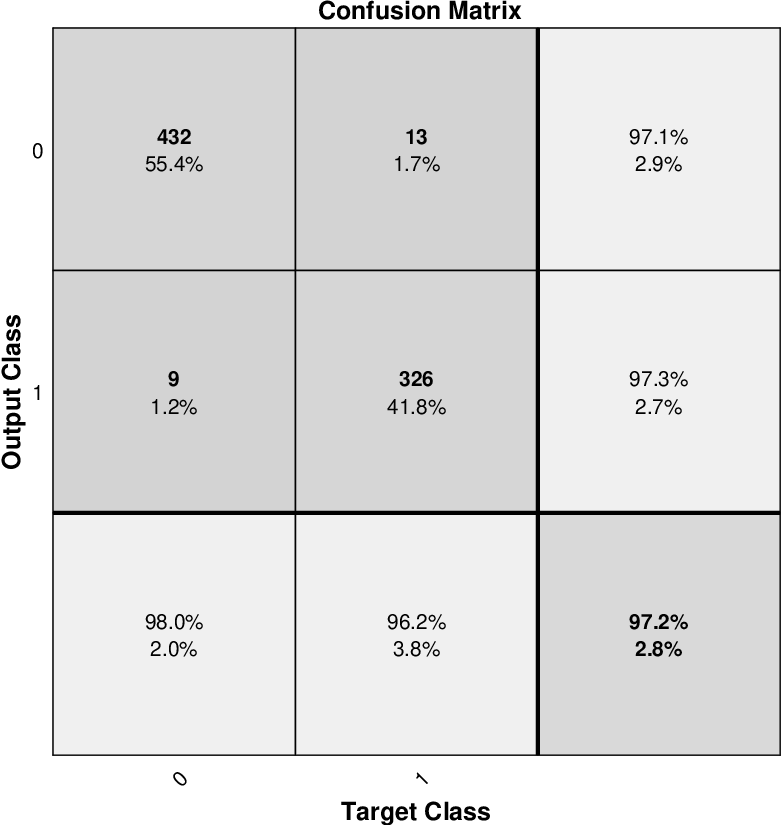}
\caption{Confusion matrix displaying classification counts for all observations. }
\label{fig:cf}
\end{figure}

\begin{figure}[!ht]
	\centering
	\includegraphics[width=180mm]{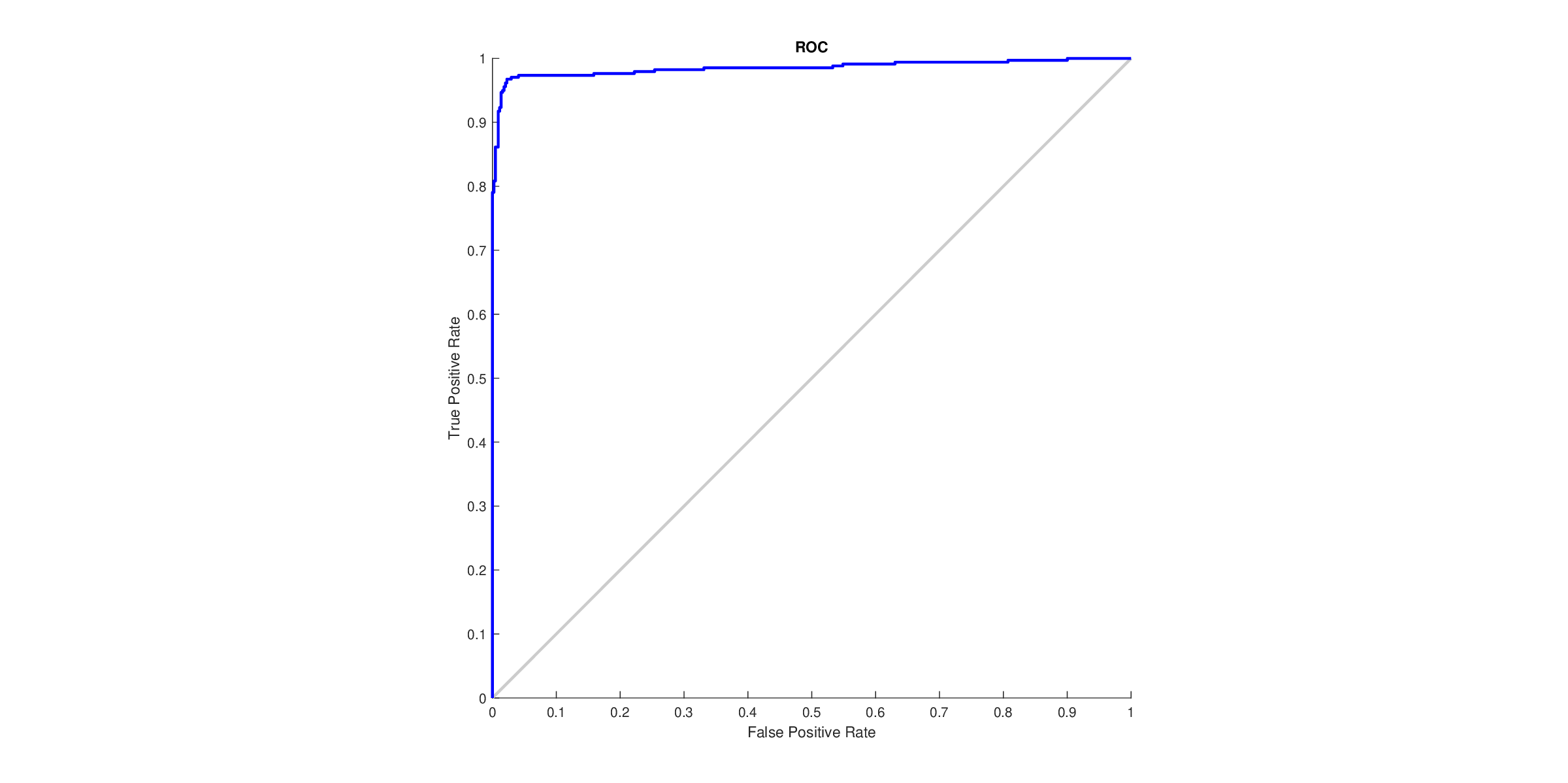}
	\caption{ROC curve for the neural network shows a good area under the curve (AUC) near $99\%$.}
	\label{fig:roc}
\end{figure}

To assess the performance of the proposed methodology, we adopted a supervised testing approach using the 780 signals described above in Section~\ref{ssec:data} for training, validation, and testing stages. 
Table~\ref{tab:perf} lists the percentage of correct classifications in terms of sensitivity, specificity, prevalence, precision, and accuracy. The abbreviations are SEN (sensitivity), SPE (specificity), PREV (prevalence), PR (precision), and ACC (accuracy). Fig.~\ref{fig:cf} shows the confusion matrix results, and Fig.~\ref{fig:roc} shows a good area under the curve (AUC) near 99\% of the ROC curve for the neural network. 

\begin{table}[!ht]
	\begin{center}
		\begin{tabular}{|c|c|c|c||}
			\hline
			&\multicolumn{3}{|c||}{\textbf{Epochs}}\\
			\hline\hline
			\textbf{Stage} & \textbf{\textit{0}}& \textbf{\textit{7}}& \textbf{\textit{13}} \\
			\hline
			Train 	& 0.256 	& 0.025 & 0.024\\
			\hline
			Test 	& 0.248	& 0.033 & 0.032 \\
			\hline
			Validation & 0.257	& 0.017 & 0.018\\
			\hline
		\end{tabular}
	\end{center}
    \vspace{0.3cm}
 \caption{Values of the error $\lambda$ for main epochs. Epoch 7 is the best performance.}
\label{tab:epochs}
\end{table}

\begin{table}[!ht]
\centering
\begin{tabular}{|c| c| c| c| c| c| c| c| c||}
\hline
    Metric & TPR & TNR & FNR & FPR & Misc & Pr & Pv & ACC\\ \hline
    Percentage & 98 & 96.2 & 2 & 3.8 & 2.8 & 98 & 56.5 &  97.2 \\
\hline	 	
\end{tabular}
\vspace{0.3cm}
\caption{Neural network performance for all signals. Percentage of good classification, in terms of TPR = True Positives Rate or \emph{Sensitivity}; TNR = True Negative Rate or \emph{specificity}; FPR = False Positive Rate; FNR = False Negative Rate; Misc = Misclassification; Pr = Precision; Pv= Prevalence and ACC = Accuracy (ACC).}
\label{tab:perf}
\end{table}

\section{Conclusion}
\label{sec:con}

This study introduced a lightweight hybrid pipeline that combines classical statistical descriptors with a shallow artificial neural network (ANN) for automatic spike– and–wave discharge (SWD) detection in long-term EEG. Using only the mean (\(\mu\)) and the standard deviation (\(\sigma\)) of a moving-average residual as input features, the system achieved promising performance on \num{780} monopolar \SI{256}{\hertz} recordings from 15 patients collected at Fundación Lucha contra las Enfermedades Neurológicas Infantiles (FLENI): sensitivity \SI{98}{\percent}, specificity \SI{96.2}{\percent}, and overall accuracy \SI{97.2}{\percent}.  These results show that interpretable, low-dimensional features can offer near-expert accuracy without the computational overhead of deep architectures.

Several limitations should be acknowledged. First, the evaluation cohort is moderate in size and drawn from a single centre, so the method’s generalisability to other clinical settings remains to be demonstrated. Second, the analysis was limited to monopolar channels; performance on multi-channel or ambulatory EEG is unknown. Third, the training time of ANN grows with the number of detected events, which may impede real-time deployment unless the network is pruned or accelerated. Finally, although more transparent than deep convolutional models, the ANN still lacks explicit uncertainty estimates.

Future work will therefore focus on multi-centre validation across diverse montages and patient groups, network-compression or incremental-learning techniques for real-time use, and the fusion of the present statistical features with complementary time–frequency or connectivity measures. A prospective clinical study is also planned to quantify the pipeline’s impact on diagnostic workload and patient outcomes.

\section*{Acknowledgment}
The authors would like to thank MD. Carlos D'Giano from FLENI, and MD. Valeria Muro from Hospital Brit\'anico for useful comments and discussions during the process of this study.

\bibliographystyle{unsrt}

\begin{thebibliography}{10}

\bibitem{Hamalainen1993}
M.~H{\"a}m{\"a}l{\"a}inen, R.~Hari, R.~J. Ilmoniemi, J.~Knuutila, and O.~V. Lounasmaa.
\newblock Magnetoencephalography: Theory, instrumentation, and applications to noninvasive studies of the working human brain.
\newblock {\em Review of Modern Physics}, 65(2):413--459, 1993.

\bibitem{Regan2010}
S.~O'Regan, S.~Faul, and W.~Marnane.
\newblock Automatic detection of {EEG} artefacts arising from head movements.
\newblock In {\em Proc.\ Annual Int.\ Conf.\ of the IEEE Engineering in Medicine and Biology Society (EMBC)}, pages 6353--6356, 2010.

\bibitem{EpilepsySpectrum2012}
M.~J. England, C.~T. Liverman, A.~M. Schultz, and L.~M. Strawbridge.
\newblock {\em Epilepsy Across the Spectrum: Promoting Health and Understanding}.
\newblock The National Academies Press, Washington, DC, 2012.

\bibitem{WHO2024Epilepsy}
{World Health Organization}.
\newblock Epilepsy: Key facts.
\newblock WHO Fact Sheet, February 2024.
\newblock Accessed 28 May 2025.

\bibitem{NiedermeyerDaSilva2010}
E.~Niedermeyer and F.~L. da~Silva.
\newblock {\em Electroencephalography: Basic Principles, Clinical Applications, and Related Fields}.
\newblock Lippincott Williams and Wilkins, 2010.

\bibitem{QuinteroRincon2021}
A.~Quintero-Rinc{\'o}n, C.~D'Giano, and H.~Batatia.
\newblock Artefacts detection in {EEG} signals.
\newblock In {\em Advances in Signal Processing: Reviews}, volume~2. 2021.

\bibitem{Markand1977}
O.~N. Markand.
\newblock Slow spike--wave activity in {EEG} and associated clinical features: Often called {L}ennox or {L}ennox--{G}astaut syndrome.
\newblock {\em Neurology}, 27(746):57, 1977.

\bibitem{Quintero2018b}
A.~Quintero-Rinc{\'o}n, M.~Alanis, V.~Muro, and C.~D'Giano.
\newblock Spike-and-{W}ave detection in epileptic signals using cross-correlation and decision trees.
\newblock {\em Revista Argentina de Bioingeniería}, 22(4):3--6, 2018.

\bibitem{Pinault2001}
D.~Pinault, M.~Vergnes, and C.~Marescaux.
\newblock Medium-voltage 5--9~hz oscillations give rise to spike-and-wave discharges in a genetic model of absence epilepsy: In~vivo dual extracellular recording of thalamic relay and reticular neurons.
\newblock {\em Neuroscience}, 105(1):181--201, 2001.

\bibitem{VanHese2009}
P.~Van~Hese, J.~Martens, L.~Waterschoot, P.~Boon, and I.~Lemahieu.
\newblock Automatic detection of spike and wave discharges in the {EEG} of genetic absence epilepsy rats from strasbourg.
\newblock {\em IEEE Transactions on Biomedical Engineering}, 56(3):706--717, 2009.

\bibitem{Ovchinnikov2010}
A.~Ovchinnikov, A.~L{\"u}ttjohann, A.~Hramov, and G.~van Luijtelaar.
\newblock An algorithm for real-time detection of spike--wave discharges in rodents.
\newblock {\em Journal of Neuroscience Methods}, 194(1):172--178, 2010.

\bibitem{Bergstrom2013}
Rachel~A Bergstrom, Jee~Hyun Choi, Armando Manduca, Hee-Sup Shin, Greg~A Worrell, and Charles~L Howe.
\newblock Automated identification of multiple seizure-related and interictal epileptiform event types in the {EEG} of mice.
\newblock {\em Scientific reports}, 3(1):1483, 2013.

\bibitem{Rodgers2015}
K.~Rodgers, F.~Dudek, and D.~Barth.
\newblock Progressive, seizure-like, spike-wave discharges are common in both injured and uninjured {S}prague--{D}awley rats: {I}mplications for the fluid percussion injury model of post-traumatic epilepsy.
\newblock {\em Journal of Neuroscience}, 35(24):9194--9204, 2015.

\bibitem{Blumenfeld2005}
H.~Blumenfeld.
\newblock Cellular and network mechanisms of spike-wave seizures.
\newblock {\em Epilepsia}, 46(9):21--33, 2005.

\bibitem{Avoli2012}
M.~Avoli.
\newblock A brief history on the oscillating roles of thalamus and cortex in absence seizures.
\newblock {\em Epilepsia}, 53(5):779--789, 2012.

\bibitem{QuinteroRincon2020a}
A.~Quintero-Rinc{\'o}n, V.~Muro, C.~D'Giano, J.~Prendes, and H.~Batatia.
\newblock Statistical model-based classification to detect patient-specific spike-and-wave in {EEG} signals.
\newblock {\em Computers}, 9(4):2--14, 2020.

\bibitem{Luttjohann2022}
A.~L{\"u}ttjohann and G.~van Luijtelaar.
\newblock Spike and wave discharges, absence epilepsy, centromedian thalamic nucleus, posterior thalamic nucleus, reticular thalamic nucleus, anterior thalamic nucleus, thalamocortical network.
\newblock {\em Epilepsy Research}, 182:106918, 2022.

\bibitem{Gobbo2021}
D.~Gobbo, A.~Scheller, and F.~Kirchhoff.
\newblock From physiology to pathology of cortico-thalamo-cortical oscillations: {A}stroglia as a target for further research.
\newblock {\em Frontiers in Neurology}, 12:661408, 2021.

\bibitem{QuinteroRincon2019d}
A.~Quintero-Rinc{\'o}n, C.~Carenzo, J.~Ems, L.~Hirschson, V.~Muro, and C.~D'Giano.
\newblock Spike-and-wave epileptiform discharge pattern detection based on {K}endall's tau-b coefficient.
\newblock {\em Applied Medical Informatics}, 41(1):1--8, 2019.

\bibitem{Samie2018}
F.~E. Abd~El-Samie, T.~N. Alotaiby, M.~I. Khalid, S.~A. Alshebeili, and S.~A. Aldosari.
\newblock A review of {EEG} and {MEG} epileptic spike detection algorithms.
\newblock {\em IEEE Access}, 6:60673--60688, 2018.

\bibitem{Webber1996}
W.~Webber, R.~Lesser, R.~Richardson, and K.~Wilson.
\newblock An approach to seizure detection using an artificial neural network.
\newblock {\em Electroencephalography and Clinical Neurophysiology}, 98:250--272, 1996.

\bibitem{Wilson2004}
S.~Wilson, M.~Scheuer, R.~Emerson, and A.~Gabor.
\newblock Seizure detection: {E}valuation of the {REVEAL} algorithm.
\newblock {\em Clinical Neurophysiology}, 115(10):2280--2291, 2004.

\bibitem{Handa2023}
P.~Handa, E.~Gupta, S.~Muskan, and N.~Goe.
\newblock A review on software and hardware developments in automatic epilepsy diagnosis using {EEG} datasets.
\newblock {\em Expert Systems}, 40(9):e13374, 2023.

\bibitem{Zhang2023}
X.~Zhang, D.~Huang, H.~Li, Y.~Zhang, Y.~Xia, and J.~Liu.
\newblock Self-training maximum classifier discrepancy for {EEG} emotion recognition.
\newblock {\em CAAI Transactions on Intelligence Technology}, 8(4):1480--1491, 2023.

\bibitem{QuinteroRincon2019b}
Bassem Bouaziz, Lotfi Chaari, Hadj Batatia, and Antonio Quintero-Rinc{\'o}n.
\newblock Epileptic seizure detection using a convolutional neural network.
\newblock In {\em Digital Health Approach for Predictive, Preventive, Personalised and Participatory Medicine}, pages 79--86. Springer, 2019.

\bibitem{Turk2019}
O.~Turk and M.~Ozerdem.
\newblock Epilepsy detection by using scalogram-based convolutional neural network from {EEG} signals.
\newblock {\em Brain Sciences}, 9(5):1--16, 2019.

\bibitem{Sriraam2018}
N.~Sriraam, S.~Raghu, K.~Tamanna, L.~Narayan, M.~Khanum, A.~S. Hegde, and A.~B. Kumar.
\newblock Automated epileptic seizures detection using multi-features and multilayer perceptron neural network.
\newblock {\em Brain Informatics}, 5(2):--, 2018.

\bibitem{Assi2018}
E.~B. Assi, L.~Gagliano, S.~Rihana, D.~Nguyen, and M.~Sawan.
\newblock Bispectrum features and multilayer perceptron classifier to enhance seizure prediction.
\newblock {\em Scientific Reports}, 8(1):15491, 2018.

\bibitem{Hussein2018}
R.~Hussein, H.~Palangi, R.~Ward, and Z.~J. Wang.
\newblock Epileptic seizure detection: {A} deep learning approach.
\newblock arXiv:\,1803.09848, 2018.
\newblock Preprint.

\bibitem{Truong2018}
N.~Truong, A.~Nguyen, L.~Kuhlmann, M.~Bonyadi, J.~Yang, S.~Ippolito, and O.~Kavehei.
\newblock Convolutional neural networks for seizure prediction using intracranial and scalp electroencephalogram.
\newblock {\em Neural Networks}, 105:104--111, 2018.

\bibitem{Acharya2018}
U.~Acharya, S.~Oh, Y.~Hagiwara, J.~Tan, and H.~Adeli.
\newblock Deep convolutional neural network for the automated detection and diagnosis of seizure using {EEG} signals.
\newblock {\em Computers in Biology and Medicine}, 100:270--278, 2018.

\bibitem{Schirrmeister2017}
R.~Schirrmeister, J.~Springenberg, L.~Fiederer, M.~Glasstetter, K.~Eggensperger, M.~Tangermann, F.~Hutter, W.~Burgard, and T.~Ball.
\newblock Deep learning with convolutional neural networks for {EEG} decoding and visualization.
\newblock {\em Human Brain Mapping}, 38(11):5391--5420, 2017.

\bibitem{Johansen2016}
A.~Johansen, J.~Jin, T.~Maszczyk, J.~Dauwels, S.~Cash, and M.~Westover.
\newblock Epileptiform spike detection via convolutional neural networks.
\newblock In {\em Proc.\ IEEE Int.\ Conf.\ on Acoustics, Speech, and Signal Processing (ICASSP)}, pages 754--758, 2016.

\bibitem{Gabor1992}
A.~Gabor and M.~Seyal.
\newblock Automated interictal {EEG} spike detection using artificial neural networks.
\newblock {\em Electroencephalography and Clinical Neurophysiology}, 83(5):271--280, 1992.

\bibitem{Jando1993}
G.~Jand{\'o}, R.~Siegel, Z.~Horv{\'a}th, and G.~Buzs{\'a}ki.
\newblock Pattern recognition of the electroencephalogram by artificial neural networks.
\newblock {\em Electroencephalography and Clinical Neurophysiology}, 86(2):100--109, 1993.

\bibitem{Gevins1986}
A.~Gevins and N.~Morgan.
\newblock Classifier-directed signal processing in brain research.
\newblock {\em IEEE Transactions on Biomedical Engineering}, BME-33(12):1054--1068, 1986.

\bibitem{Guo2010}
L.~Guo, D.~Rivero, J.~Dorado, J.~R.~nal, and A.~Pazos.
\newblock Automatic epileptic seizure detection in {EEG}s based on line length feature and artificial neural networks.
\newblock {\em Journal of Neuroscience Methods}, 194(1):88--96, 2010.

\bibitem{Dhif2017}
I.~Dhif, K.~Hachicha, A.~Pinna, S.~Hochberg, I.~Mhedhbi, and P.~Garda.
\newblock Epileptic seizure detection based on expected activity measurement and neural network classification.
\newblock In {\em Proc.\ Annual Int.\ Conf.\ of the IEEE Engineering in Medicine and Biology Society (EMBC)}, pages 2814--2817, 2017.

\bibitem{Aguiar2015}
K.~de~Aguiar, F.~M.~G. Franca, V.~C. Barbosa, and C.~A.~D. Teixeira.
\newblock Early detection of epilepsy seizures based on a weightless neural network.
\newblock In {\em Proc.\ 37th Annual Int.\ Conf.\ of the IEEE Engineering in Medicine and Biology Society (EMBC)}, pages 4470--4474, 2015.

\bibitem{Gabor1996}
A.~Gabor, R.~Leach, and F.~Dowla.
\newblock Automated seizure detection using a self-organising neural network.
\newblock {\em Electroencephalography and Clinical Neurophysiology}, 89(1):257--266, 1996.

\bibitem{Ponce2016}
M.~Alfaro-Ponce, A.~Arguelles, and I.~Chairez.
\newblock Pattern recognition for electroencephalographic signals based on continuous neural networks.
\newblock {\em Neural Networks}, 79:88--96, 2016.

\bibitem{Hyndman2021}
R.~J. Hyndman and G.~Athanasopoulos.
\newblock {\em Forecasting: Principles and Practice}.
\newblock OTexts, 2021.

\bibitem{Thomopoulos2017}
N.~T. Thomopoulos.
\newblock {\em Statistical Distributions: Applications and Parameter Estimates}.
\newblock Springer, 2017.

\bibitem{NNDesign2014}
M.~T. Hagan, H.~B. Demuth, M.~H. Beale, and O.~D. Jes{\'u}s.
\newblock {\em Neural Network Design}.
\newblock Martin Hagan, 2014.

\bibitem{QuinteroRincon2018a}
A.~Quintero-Rinc{\'o}n, M.~Pereyra, C.~D'Giano, M.~Risk, and H.~Batatia.
\newblock Fast statistical model-based classification of epileptic {EEG} signals.
\newblock {\em Biocybernetics and Biomedical Engineering}, 38(4):877--889, 2018.

\bibitem{MasinoQuintero2024}
Nicolás Masino and Antonio Quintero-Rincón.
\newblock Effect sizes as a statistical feature-selector-based learning to detect breast cancer.
\newblock In {\em 2024 IEEE Biennial Congress of Argentina (ARGENCON)}, pages 1--7, 2024.

\bibitem{Masino_Quintero-Rincon_2025}
Nicolas~Martín Masino and Antonio Quintero-Rincón.
\newblock Diagnóstico de cáncer de mama usando el tamaño del efecto d de cohen como selector de características.
\newblock {\em Inteligencia Artificial}, 28(75):260--280, 2025.

\bibitem{Twomey1995}
J.~M. Twomey and A.~E. Smith.
\newblock Performance measures, consistency and power for artificial neural network models.
\newblock {\em Mathematical and Computer Modelling}, 21(1--2):243--258, 1995.

\end{thebibliography}

\end{document}